\documentclass[conference]{IEEEtran}
\IEEEoverridecommandlockouts
\usepackage[boxruled]{algorithm2e}
\usepackage{enumerate}
\usepackage{multirow,array}
\usepackage{dirtytalk}
\usepackage{cite}
\usepackage{graphicx}
\usepackage{psfrag}
\usepackage{caption}
\usepackage{subcaption}
\usepackage{url}
\usepackage{amsmath}
\usepackage{amsthm}
\usepackage{kbordermatrix}
\usepackage{array}
\usepackage{algorithm2e}
\usepackage{amssymb}
\usepackage{amsfonts}
\usepackage{float}
\usepackage{textcomp}
\usepackage{xcolor}
\usepackage{tabu}
\usepackage{array}
\newcolumntype{P}[1]{>{\centering\arraybackslash}p{#1}}
\newcolumntype{M}[1]{>{\centering\arraybackslash}m{#1}}

\def\BibTeX{{\rm B\kern-.05em{\sc i\kern-.025em b}\kern-.08em
    T\kern-.1667em\lower.7ex\hbox{E}\kern-.125emX}}
\usepackage{float}

\renewcommand{\boxed}[1]{\text{\fboxsep=.2em\fbox{\m@th$\displaystyle#1$}}}

\newtheorem*{conjecture*}{Conjecture}
\newtheorem{lemma}{Lemma}

\newtheorem{remark}{Remark}
\newtheorem{definition}{Definition}
\newtheorem{theorem}{Theorem}
\newtheorem{example}{Example}

\title{Coded Caching based on Combinatorial Designs}
\begin{document}

\author{

  \IEEEauthorblockN{Shailja Agrawal, K V Sushena Sree, Prasad Krishnan}
  \IEEEauthorblockA{
                    International Institute of Information Technology, Hyderabad\\ 
                    Email: \{shailja.agrawal@research. , sushena.sree@research. , prasad.krishnan@\}iiit.ac.in}

\vspace{-0.2cm}
}

\maketitle
\begin{abstract}
We consider the standard broadcast setup with a single server broadcasting information to a number of clients, each of which contains local storage (called \textit{cache}) of some size, which can store some parts of the available files at the server. The centralized coded caching framework, consists of a caching phase and a delivery phase, both of which are carefully designed in order to use the cache and the channel together optimally. In prior literature, various combinatorial structures have been used to construct coded caching schemes. In this work, we propose a binary matrix model to construct the coded caching scheme. The ones in such a \textit{caching matrix} indicate uncached subfiles at the users. Identity submatrices of the caching matrix represent transmissions in the delivery phase. Using this model, we then propose several novel constructions for coded caching based on the various types of combinatorial designs. While most of the schemes constructed in this work (based on existing designs) have a high cache requirement (uncached fraction being $\Theta(\frac{1}{\sqrt{K}})$ or $\Theta(\frac{1}{K})$, $K$ being the number of users), they provide a rate that is either constant or decreasing ($O(\frac{1}{K})$) with increasing $K$, and moreover  require competitively small levels of subpacketization (being $O(K^i), 1\leq i\leq 3$),  which is an extremely important parameter in practical applications of coded caching.  We mark this work as another attempt to exploit the well-developed theory of combinatorial designs for the problem of constructing caching schemes, utilizing the binary caching model we develop.
\end{abstract}

\section{Introduction}
The increase in wireless data traffic necessitates the role of broadcast communication, where a single server is delivering information payloads to multiple clients at the same time. The network coding paradigm is especially useful in this regard, enabling the coding of information intended for multiple clients and delivering the coded information at once to all of them. Coded caching, which exploits network coding \cite{NC} for the specific broadcast setting where each of the clients have local storage (called \textit{cache}), was proposed  in \cite{MaN} and has emerged as a valuable technique to use the communication channel efficiently. 

The coded caching scenario as in \cite{MaN}, consists of clients indexed by some set ${\cal K}$ of size $K$ and possessing some cache, connected to a single server through an error free shared link. The library of files at the server consists of $N$ files of same size, which are denoted as $W_i : \forall i \in [N]$. Each file consists of $F$ non-overlapping subfiles of same size, where $F$ is known as the \textit{subpacketization level}. The subfiles of $W_i$ are labelled as $W_{i,f}: f\in {\cal F}$, ($W_{i,f}$ is assumed to take values from abelian group) where ${\cal F}$ is a set of size $F$. The centralized coded caching framework consists of two phases : the \textit{placement phase} and the \textit{delivery phase}. The placement phase occurs during non-peak hours. In the placement phase, the communication channel is utilized so that each client stores some $\frac{M}{N}$ fraction of each file in the library in its cache, where $M$ reflects the cache size. The delivery phase corresponds to peak-hours. In the delivery phase (during peak hours), the demands of the users pop up. In the coded caching paradigm, the server broadcasts \textit{coded} transmissions such that the demands of all the users are satisfied. As in \cite{MaN}, the rate $R$ of the coded caching scheme is defined as the ratio of the number of bits transmitted to the size of each file, and can be calculated as 
\[
\text{Rate}~R = \small \frac{\text{Number of transmissions in the delivery phase}}{\text{Subpacketization level}},
\]
when each transmission is of the same size as any subfile.


Though the coded caching framework presented in \cite{MaN} achieves an optimal rate, its exponential increase in subpacketization with respect to the number of users at constant $\frac{M}{N}$ is a major setback for its practical implementation. The scheme presented in \cite{OPDA} gives reduced subpacketization by using a combinatorial structure that designs both the placement and the delivery phase together, called as the Placement Delivery Array (PDA). This idea has been further extended in \cite{PDA} by using strong edge coloring of an associated bipartite graph.  In \cite{RM}, resolvable designs derived from linear block codes have been used to reduce subpacketization. All of these schemes offered reductions in subpacketization as compared to \cite{MaN}, at the cost of some increase in the rate, for constant memory fraction $\frac{M}{N}$. However, to the best of our knowledge, most of the schemes (for reasonable values of $K$) available in literature require subpacketization exponential in $K$. A subpacketization subexponential in $K$ has been obtained in \cite{PK} using a line graph model for coded caching along with a projective geometry based scheme. For constant rate, the scheme in \cite{PK} achieves a subpacketization level of $O(q^{(log_qK)^2})$, however demanding that the uncached fraction, $(1-\frac{M}{N})=\Theta(\frac{1}{\sqrt{K}})$. 

 The first contribution of this work is to present a new binary matrix model for coded caching. In Section \ref{model}, we introduce the concept of using a constant row-weight binary matrix for describing the coded caching scheme. We call these as \textit{caching matrices}. The 1s in the binary matrix indicate uncached subfiles in the users. Identity submatrices of the caching matrix correspond to transmissions which enable the clients (involved in any transmission) to decode precisely one missing subfile each from that transmission. Thus, `covering'  the 1s in the caching matrix using identity submatrices provides a valid delivery scheme. The framework we present using binary matrices are closely (and obviously, as the reader shall see) related to the PDA schemes. However the advantage is that this viewpoint opens up a much larger space, viz. the space of all constant row-weight binary matrices, for searching for good caching schemes.  

Following this, we use the binary matrix model for constructing novel caching schemes derived from a variety of combinatorial designs. Towards that end, Section \ref{designs} describes important terminologies related to combinatorial designs. In Sections \ref{BIBD}-\ref{transvelsal}, we elaborate on the construction of caching matrices using different combinatorial designs. In particular, we employ Steiner systems ($t$-designs with special properties), balanced incomplete block designs, and transversal designs, to construct caching matrices. When we employ existing designs from combinatorics literature to these constructions, the caching schemes which we get demand a low uncached fraction, i.e., $1-\frac{M}{N}=\Theta(\frac{1}{K^i}), i=\frac{1}{2},1$. This is a disadvantage. However this disadvantage is traded off by a deep reduction in the rate as well as the subpacketization levels, with the schemes achieving a constant rate or even a rate that is $O(\frac{1}{K})$, with subpacketization levels being  $O(K^i), 1\leq i\leq 3$. Section \ref{summary} summarizes all our constructions and discusses the asymptotics of each. We end the paper in Section \ref{discussion} with some promising directions that can possibly help us to remedy the issue of high cache requirements at the users.


\textit{Notations and Terminology: }
For any positive integer $N$, we denote by $[N]$ the set $\{1,\hdots,N\}$. For a set ${\cal X}$ and some positive integer $t\leq {|\cal X|}$, we denote the set of all $t$-sized subsets of ${\cal X}$ by $\binom{{\cal X}}{t}$. For a matrix $A$ whose rows are indexed by a finite set ${\cal R}$ and columns are indexed by a finite set ${\cal C}$, the element in the $r^{th}$ row  $(r \in {\cal R})$ and $l^{th}$ column  $(l \in {\cal C})$ is denoted as $A(r,l)$. For sets $A,B,$ $A\backslash B$ denotes the elements in $A$ but not in $B$. For some element $i$, we also denote $A\backslash \{i\}$ by $A\backslash i$. For $j\in\{0,1,\hdots,(k-1)\}$, we denote $(j+1)mod~k$ by $(j {\oplus_k}1)$.



\section{A binary matrix model for coded caching}
\label{model}
In this section we describe how a coded caching scheme can be derived from a binary matrix with constant row weight. 
\begin{definition}
[Caching Matrix] Consider a matrix $C$ with entries from $\{0,1\}$ with rows indexed by a $K$-sized set ${\cal U}$ and columns indexed by a $F$-sized set ${\cal F}$  such that the number of $1$'s in each row is constant (say $Q$).
Then the matrix $C$ defines a caching scheme with $K$ users (indexed by ${\cal U}$), subpacketization $F$ (indexed by ${\cal F}$) and $(1-\frac{M}{N}) = \frac{Q}{F}$ as follows:
\begin{itemize}
\item User $u\in {\cal U}$ caches $W_{i,f}: \forall i \in [N]$ if $C(u,f) = 0$ and does not cache it if $C(u,f) = 1$.
\end{itemize}
We then call the matrix $C$ as a $({\cal U}, {\cal F}, (1-\frac{M}{N}))$ - \textit Caching Matrix.
\end{definition}

A subfile $W_{i,f}$ is said to be \textit{missing} at a user $u$ if it is not available at its cache. The demand of a user $u$ in the delivery phase is denoted by $W_{{d_u}}$ for some $d_u \in [N]$.  In order to construct a transmission scheme, we first describe one transmission based on the above described matrix based caching scheme, which will serve a number of users. 
Note that a submatrix of $C$ can be specified by a subset of the row indices ${\cal U}$ and a subset of column indices ${\cal F}$. We now define an \textit{identity submatrix} $C_i$ of matrix $C$.

\begin{definition}
[Identity Submatrix] An $l \times l$ submatrix $C_i$ of the matrix $C$ is an identity submatrix of size $l$ if its columns correspond to the identity matrix of size $l$ permuted in some way.
\end{definition}


\begin{lemma}
\label{transmissions_def}
Consider an identity submatrix of $C$ given by rows $\{u_1,u_2,..,u_l: u_i \in {\cal U}\}$ and columns $\{f_1,f_2,..,f_l: f_i \in {\cal F}\}$, such that $C(u_i,f_i)=1, \forall i \in [l]$, while $C(u_i,f_j)=0, \forall i,j \in [l]$ where $i\neq j$. For each $i \in [l]$, the subfile $W_{d_{u_i},f_i}$ is not available at user $u_i$ and can be decoded from the transmission $\sum_{i=1}^l W_{d_{u_i},f_i}$.
\end{lemma}

\begin{IEEEproof}
By definition of identity submatrix, for each $i \in [l]$ the subfile $W_{d_{u_i},f_i}$ is not available at user $u_i$ but is available at the users $\{u_1,u_2,..,u_l\} \backslash u_i$. Hence each user $u_i : i \in [l]$ can decode the subfile $W_{d_{u_i},f_i}$ which is not available at its cache from the transmission $\sum_{i=1}^l W_{d_{u_i},f_i}$.
\end{IEEEproof}
We shall use Lemma \ref{transmissions_def} to describe the complete transmission scheme. For that purpose we introduce few more terminologies.

For a \textit {caching matrix} $C$, suppose $C(u,f) =1$ for some $u \in {\cal U}$ and $f \in {\cal F}$. The entry $C(u,f) =1$ is said to be \textit {covered} by the identity submatrix $B$ if $u$ and $f$ correspond to some row and column index of $B$ respectively.
\begin{definition}[Identity Submatrix Cover]
Consider a set $\mathfrak {C}=\{C_1,...,C_S\}$ consisting of $S$ identity submatrices of a caching matrix $C$ such that any $C(u,f) =1$ in $C$ is \textit {covered} by atleast one $C_i$ such that $i\in S$. Then, $\mathfrak {C}$ is called an Identity Submatrix Cover of $C$.

\end{definition}
 We now describe how an identity submatrix cover is used to form a transmission scheme.
\begin{theorem}
\label{rate def}
Consider an identity submatrix cover $\mathfrak{C}= \{C_1,C_2,..,C_S\}$ of a caching matrix $C$. Then the transmission corresponding to $C_i:i \in [S]$ according to Lemma \ref{transmissions_def}, is a valid transmission scheme (i.e the scheme  satisfies all the user demands) for the caching scheme defined by $C$ and the rate of the transmission scheme, $R = \frac{S}{F}$.

\end{theorem}
\begin{IEEEproof}
Pick some arbitrary missing subfile $ W_{d_{u},f}$ of user $u$. Then $C(u,f) =1$ and this entry of $C$ will be covered by atleast one of the identity submatrices, say $C_i$ in $\mathfrak{C}$ since $\mathfrak{C}$ is an identity submatrix cover of $C$. The transmission corresponding to the identity submatrix $C_i$ given by Lemma \ref{transmissions_def} will ensure that the subfile $ W_{d_u,f}$ will be decoded by the corresponding user $u$ where it is missing. Hence, the transmissions corresponding to $C_i \in \mathfrak{C}$ enables decoding of any arbitrary missing subfile. Since the number of identity submatrices  in $\mathfrak{C}$ is $S$, the rate of the transmission scheme is, $R=\frac{S}{F}$.
\end{IEEEproof}

We also need the idea of an \textit{overlap} between identity submatrices of $C$, which enables us to prove some results in this paper. 
\begin{definition}
[Overlap] An \textit{overlap} between identity submatrices occurs when some entry $C(u,f)=1$ in matrix $C$ is covered by more than one identity submatrix of $C$.
\end{definition}

\section{introduction to Combinatorial designs}
\label{designs}
In the previous section, we have developed a binary matrix model for the caching problem. In sections \ref{BIBD}-\ref{transvelsal}, we will use combinatorial designs to construct caching matrices. For that purpose we first review some of the basic definitions related to designs and their constructions. For more details reader is referred to \cite{Drs}\cite{col}.
\begin{definition}[Design $({\cal X},{\cal A})$] A design is a pair $({\cal X},{\cal A})$ such that the following properties are
satisfied:
\newline (D1). ${\cal X}$ is a set of elements called points, and
\newline   (D2). ${\cal A}$ is a collection (i.e., multiset) of nonempty subsets of ${\cal X}$ called blocks.
\end{definition}
We now define $t$-designs.

\begin{definition}[$t$-designs]
Let $v, k, \lambda ,$ and $t$ be positive integers such that $v>k \geq t$. A $t$-$(v, k, \lambda)$-design (or simply $t$-design) is a design $(\cal X, A)$ such that the following properties are satisfied: \newline (T1). ${|\cal X|} = v$,
 \newline   (T2). Each block contains exactly $k$ points, and
  \newline   (T3). Every set of $t$ distinct points is contained in exactly $\lambda$ blocks.
  \end{definition}
Consider a nonempty $Y\subseteq {\cal X}$ such that $|Y| = s \leq t$. Then there are exactly
\begin{align}
\label{s_points}
{\lambda_s}  = \lambda \frac{\binom{v-s}{t-s}}{\binom{k-s}{t-s}}
\end{align}
blocks in ${\cal A}$ that contain all the points in $Y$. It can also be shown that $b={\lambda_0}=\lambda \dfrac{\binom{v}{t}}{\binom{k}{t}}$ is the number of blocks in $t$-designs. 

\begin{example}
\label{t_ex}
[Parametrized Constructions]
A $t$-design with $\lambda =1$ (i.e  $t$-$(v,k,1)$ design) is called a Steiner system and its existence is discussed in \cite{keevash}. A construction of Steiner system for $t=3$ and $t=4$ is presented in \cite{pappu}. Other general constructions for Steiner systems can be found in \cite{col}. Here we use a specific construction.

\begin{itemize}
    \item A construction of Steiner system with parameters $t=3,~ v=q^{2}+1,~ k=q+1$ is presented in \cite{Drs}, where $q$ is a prime power such that $q \geq 2$.
\end{itemize}
\end{example}

In the following examples and some others in this paper, we drop the parentheses and the commas in writing the blocks explicitly (for instance block $\left\{l,m,n \right\}$ is written as $lmn$).
\begin{example}
\label{ex4}
A 3-(8,4,1) design (Steiner system)
\newline ${\cal X} = \left\{1,2,3,4,5,6,7,8 \right\}$
\newline ${\cal A} = \{ 1256,3478,1357,2468,1458,2367,
1234,5678,1278,$\newline $3456,1368,2457,1467,2358 \}$.  
\end{example}

\begin{definition}[Balanced Incomplete Block Design] 
$t$-Designs with $t=2$ are called Balanced Incomplete Block Designs, (BIBD) denoted as ($v$, $k$, $\lambda$)-BIBD.
\end{definition}

By (\ref{s_points}) it follows that every ($v$, $k$, $\lambda$)-BIBD has exactly $b = {\frac {vr}{k}}$ blocks and, every point occurs in exactly 
\begin{align}
\label{r_points}
r = {\frac {\lambda(v-1)}{k-1}}
\end{align}
blocks.

\begin{example}
\label{ex1}
A (9,3,1)-BIBD.
\newline ${\cal X} = \left\{1,2,3,4,5,6,7,8,9 \right\}$
\newline ${\cal A} = \{ 357,123,456,789,147,258,369,159,267,348,168,
$\newline$ 249 \}$. The number of blocks are $b=12$ and each element in ${\cal X}$ occurs exactly in $r=4$ blocks.
\end{example}
We now define Symmetric BIBD.

\begin{definition}[Symmetric BIBD]
A $(v,k,\lambda)$-BIBD in which $b = v$ (or, equivalently, $r = k$) is called a symmetric BIBD.
\end{definition}


 
 There are only finitely many nontrivial symmetric BIBDs with $\lambda=2$ that are known \cite{sym}. As stated in \cite{Drs}, for any two blocks $A_1, A_2 \in {\cal A}$ in a symmetric BIBD 
 \begin{align}
\label{lambda}
|A_1 \cap A_2| = \lambda.
\end{align}

\begin{example}
\label{ex2}A symmetric $(11,5,2)$-BIBD
\newline ${\cal X} = \left\{1,2,3,4,5,6,7,8,9,a,b \right\}$
              \newline ${\cal A} = \{ 1689a,13467,1249b,1235a,23789,348ab,4579a, $ \newline $ 267ab, 1578b,24568,3569b \}$

\end{example}
\begin{example}
\label{bibd_const}
[Parametrized Constructions]
Some constructions of BIBD known in literature are given below:
\begin{itemize}
    \item Symmetric BIBDs with parameters $v= n^{2}+n+1$, $k= n+1$, $\lambda = 1$ are constructed in \cite{Drs} using a projective plane of order $n$, where $n$ is a prime power such that $n \geq 2$.
    
    \item BIBDs with parameters $v= n^{2}$, $k= n$, $\lambda = 1$ are constructed in  \cite{Drs} using an affine plane of order $n$ where, $n$ is a prime power such that $n \geq 2$.
    
    \item A construction of symmetric-BIBDs using affine resolvable BIBDs is presented in \cite{arbibd}.
\end{itemize}
\end{example}

We now define Transversal Designs.

\begin{definition}[Transversal Designs] A transversal design of order or groupsize $n$, blocksize $k$, and index $\lambda$, denoted as $TD_{\lambda}(k, n)$, is a triple $( \cal X, \cal G, \cal B)$, where \newline (TD1). $ {\cal X}$ is a set of $kn$ elements.
\newline     (TD2).  $\cal G$ is a partition of ${\cal X}$ into $k$ sets (the groups), each of size $n$.
\newline     (TD3).  $\cal B$ is a collection of $k$-sized subsets of $ {\cal X}$ (the blocks).
\newline     (TD4).  Every pair of elements from ${\cal X}$ is contained either in exactly one group or in exactly $\lambda$ blocks, but not both.
\end{definition}
Transversal designs in which $\lambda = 1$, are denoted by TD$(k, n)$. From the above properties, we see that  $|\cal B|$= $n^{2}$ and each element of $ {\cal X}$ occurs in $n$ blocks for $\lambda=1$ \cite{han}.
\begin{example}
\label{ex5}
A TD$(4,3)$ design is given as follows.
\newline${\cal X} =\{1,2,3,4,5,6,7,8,9,10,11,12   \} $
\newline ${\cal G} = \{ \{1,2,3 \},\{4,5,6 \}, \{7,8,9 \},\{10,11,12 \} \}$
\newline ${\cal B} = \{ \left\{1,4,7,10 \right\}, \left\{1,5,8,11 \right\}, \left\{1,6,9,12 \right\}, \left\{2,4,9,11 \right\},$ \newline $ 
\left\{2,5,7,12 \right\}, \left\{2,6,8,10 \right\}, \left\{3,4,8,12 \right\}, \left\{3,5,9,10 \right\},$ \newline $  \left\{3,6,7,11 \right\} \}$.
\end{example}
\begin{example}
\label{ex_td1}
[Parametrized Constructions] Some constructions of TD$(k,n)$ known in literature \cite{Drs} are as follows:
\begin{itemize}
    \item A transversal design with parameters $\lambda =1, k=q, n=q$ can be constructed using orthogonal arrays, where $q$ is a prime power such that $q\geq 2$.
     \item A transversal design with parameters $\lambda =1, k=q+1, n=q$ can be constructed using orthogonal arrays, where $q$ is a prime power such that $q\geq 2$.
     
\end{itemize}
\end{example}
We will use in some constructions the following idea of the \textit{incidence matrix} of a design.
\begin{definition}[Incidence Matrix] Let ($\cal X, \cal A $) be a design where ${\cal X}$ = $\left\{x_1,...,x_v \right\}$ and $\cal A$ =
$\left\{A_1,...,A_b \right\}$. The incidence matrix of ($\cal X, \cal A $) is the $v\times b$ binary matrix $M =(M(i,j))$ defined by the rule

$M(i,j) =
    \begin{cases}
      1, & \text{if}\ x_i\in A_j , \\
      0, & \text{if}\ x_i\not\in A_j. \\
    \end{cases}$
\end{definition}
\begin{example}
\label{ex_5}
Incidence Matrix for $(7,3,1)$-BIBD is given below
\newline ${\cal X} = \left\{1,2,3,4,5,6,7 \right\}$
\newline ${\cal A} = \{127,145,136,467,256,357,234 \}$
\renewcommand{\kbldelim}{(}
\renewcommand{\kbrdelim}{)}
\[
  \text{$C$} = \kbordermatrix{
    & 127 & 145 & 136 & 467 & 256 & 357 & 234 \\
    1 & 1 & 1 & 1 & 0 & 0 & 0 & 0\\
    2 & 1 & 0 & 0 & 0 & 1 & 0 & 1\\
    3 & 0 & 0 & 1 & 0 & 0 & 1 & 1\\
    4 & 0 & 1 & 0 & 1 & 0 & 0 & 1\\
    5 & 0 & 1 & 0 & 0 & 1 & 1 & 0\\
    6 & 0 & 0 & 1 & 1 & 1 & 0 & 0\\
    7 & 1 & 0 & 0 & 1 & 0 & 1 & 0
  }
\]

\end{example}

\section{Summary of Results}
\label{summary}
Table \ref{table} summarizes all the caching parameters related to the coded caching schemes to be constructed from the various designs in the forthcoming sections. The parameters are based on those of the designs using which they are constructed. Applying the results of Table \ref{table} to the parameterized constructions of designs as given in Section \ref{designs}, we get the following results.
\begin{table}[ht]

\tabulinesep=1.0mm
\centering
\captionof{table}{Parameters of Coded Caching Scheme based on Combinatorial Designs}
\begin{tabular}{ |M{1.7cm}|M {1.5cm}|M{0.7cm}|M{0.9cm}|M{1.3cm}|  }
 \hline
 
 \textbf{Combinatorial Designs} & $\left( \mathbf{1-\frac{M}{N}} \right)$ & $\mathbf{K}$ & $\mathbf{F}$ &$\mathbf{R}$     \\
 \hline

\textit{BIBD (${\lambda=1}$) {Section \ref{BIBD}}}&$\frac{k}{v}$&$v$&$\frac{v(v-1)}{k(k-1)}$&$\frac{k(k-1)}{v-1}$\\
\hline

\textit{Symmetric BIBD (${\lambda=2}$) {Section \ref{Symm_BIBD}}}&$\frac{k-1}{v}$&$v$&$kv$&$1$\\
\hline

${t}$-\textit{design (${\lambda=1}$) (Scheme $1$) {Section \ref{t_Scheme1}}} &$\frac{{\binom{k}{t}(v-t+1)}}{{\binom{v}{t}}k}$&$\binom{v}{t-1}$&$\frac{{\binom{v}{t}k}}{\binom{k}{t}}$&$\frac{{\binom{k-1}{t-1}}{\binom{k}{t}}v}{\binom{v}{t}k}$\\
\hline

${t}$-\textit{design (${\lambda=1}$) (Scheme $2$) {Section \ref{t_Scheme2}}} &$\frac{t}{v}$&$v$&$\binom{v}{t}$&$\frac{t}{v-t+1}$\\
 \hline

\textit{ Transversal Design (${\lambda=1}$) {Section \ref{transvelsal}}}&$\frac{1}{n}$&$n^{2}$&$kn$&$1$\\
\hline
\end{tabular}
\label{table}
\end{table}
\subsection{Specific Constructions}
\subsubsection{BIBDs}
The parameters of the transmission scheme (described in Section \ref{BIBD}) for the constructions described in Example \ref{bibd_const} are as follows:
\begin{itemize}
  \item Symmetric BIBDs with parameters $v= n^{2}+n+1$, $k= n+1$, $\lambda = 1$ will give a coded caching scheme with parameters $F=n^{2}+n+1$, $K=n^{2}+n+1$, Rate $= 1$, $(1-\frac{M}{N}) = \frac{n+1}{n^{2}+n+1}$.
    
    \item BIBDs with parameters $v= n^{2}$, $k= n$, $\lambda = 1$ will give a coded caching scheme with parameters $F=n^{2}+n$, $K=n^{2}$, Rate $= \frac{n}{n+1}$, $(1-\frac{M}{N}) = \frac{1}{n}$.
\end{itemize}

Note that for the above two schemes, we have $F=O(K)$, and $1-\frac{M}{N}=\Theta(\frac{1}{\sqrt{K}}),$ while $R\leq 1.$
\subsubsection{Steiner systems}
For the constructions described in Example \ref{t_ex}, the parameters of the transmission scheme presented in Section \ref{t-section} are as follows:
\paragraph{Scheme $1$}
$t$-designs with parameters $t=3,~v=q^{2}+1,~k=q+1, ~\lambda=1$,  will give a coded caching scheme with parameters $F=(q^{2}+1)(q+1)$, $K=\frac{(q^{2}+1)q^{2}}{2}$, Rate $= \frac{(q-1)}{2(q+1)}$, $(1-\frac{M}{N}) = \frac{(q-1)}{q(q^{2}+1)}$.

We note that for the above construction of the coded caching scheme, we have $F=O(K^{\frac{3}{4}}),$ $1-\frac{M}{N}=\Theta(\frac{1}{\sqrt{K}})$ and $R\leq 1.$

\paragraph{Scheme $2$}

     $t$-designs with parameters $t=3, ~v=q^{2}+1,~ k=q+1,~ \lambda=1$,  will give a coded caching scheme with parameters $F=\binom{q^{2}+1}{3}$, $K=q^{2}+1$, Rate $= \frac{3}{q^{2}-1}$, $(1-\frac{M}{N}) = \frac{3}{q^{2}+1}$.

Observe that for this scheme, the parameters of the coded caching scheme $F=O(K^3), \left(1-\frac{M}{N}\right)=\Theta(\frac{1}{K})$, while rate is also $O(\frac{1}{K}).$
\subsubsection{Transversal Designs}
The parameters of the transmission scheme (presented in Section \ref{transvelsal}) for the constructions described in Example \ref{ex_td1} are as follows:

\begin{itemize}

     \item A transversal design with parameters $\lambda =1,~ k=q,~ n=q$, will give a coded caching scheme with parameters $F=q^{2}$, $K=q^{2}$, Rate $= 1$, $(1-\frac{M}{N}) = \frac{1}{q}$.
     
     \item A transversal design with parameters $\lambda =1,~ k=q+1,~ n=q$, will give a coded caching scheme with parameters $F=q^{2}+q$, $K=q^{2}$, Rate $= 1$, $(1-\frac{M}{N}) = \frac{1}{q}$.
\end{itemize}

For both of the above constructions, we have $F=O(K)$ and $(1-\frac{M}{N})=O(\frac{1}{\sqrt{K}}).$

In the forthcoming sections, we provide constructions for caching schemes based on the above combinatorial designs. Each construction is contingent on the existence of the design of the considered type. In each such case, we define the caching matrix using the given design and obtain its parameters $K, F,$ and $(1-\frac{M}{N})$ based on the design parameters. We then define and prove an identity submatrix cover of the caching matrix based on the properties of the design. The format for the proof of the identity submatrix cover in each of the following constructions is the same, which we describe sequentially as follows.
\begin{enumerate}
    \item We describe a method to pick a submatrix of the caching matrix,  which we prove to be an identity submatrix in the following way.
    \begin{enumerate}
        \item We show that the submatrix has equal number of rows and columns. 
        \item We then show that each row and column of the submatrix has weight one. 
    \end{enumerate}
    \item We then show that the identity submatrices picked have no overlaps. 
    \item Finally we show that all the $1$'s of the caching matrix are covered by the collection of identity submatrices, thus proving that the collection forms an identity submatrix cover. 
\end{enumerate}

\section{BIBD (with $\lambda =1$) based Coded Caching Scheme}
\label{BIBD}
Consider a ($v,k,1$)-BIBD $({\cal X,A})$. We order the elements in ${\cal X}$ in some arbitrary way. For distinct $x,y \in {\cal X}$, we say that $x < y$ if $x$ comes before $y$ in the ordering of ${\cal X}$. Let the elements in block $A\in{\cal A}$ be denoted as $\{A(0),A(1),\hdots,A(k-1)\}$ where $A(0)<A(0)<..<A(k-1)$. Let $C$ denote the incidence matrix of ($v,k,1$)-BIBD.

\begin{remark}
Note that each element in ${\cal X}$ occurs in $r$ blocks. Thus each row of the incidence matrix $C$ has weight $r$. Also, each block $A\in {\cal A}$ is of size $k$, thus column weight of the matrix $C$ is $k$. Hence the binary caching matrix constructed from BIBD has constant row and column weight.
\end{remark}

Now, $(1-\frac{M}{N}) = \frac{r}{b} = \frac{k}{v}$ ($\because vr=bk$). $F =b= \frac{v(v-1)}{k(k-1)}$.  Hence, the incidence matrix $C$ of design $({\cal X,A})$ will give a $\left(v,\frac{v(v-1)}{k(k-1)},\frac{k}{v}\right)$-caching matrix.
\par For $x \in {\cal X}$, let $B_x \triangleq\left\{A \in {\cal A} : x \in A\right\}$. Note that $|B_x| = r$, by the property of $\lambda=1$ BIBD. In the next lemma, we describe a single identity submatrix of matrix $C$. 


\begin{lemma}
\label{Cx_identity}
For any $x\in {\cal X}$, let us denote $B_x$ as $B_x=\left\{A_{1},...,A_{r}\right\}$ where $x=A_{i}(j_i) : j_i\in\{0,1,\hdots,k-1\}$ i.e $x$ is the $j_i^{th}$ element in the block $A_{i}$. Consider the submatrix $C_x$ of $C$ whose columns are indexed by $B_x$ and rows are indexed by $\left\{ A_{i}(j_{i} {\oplus_k} 1)  : i\in[r], A_{i} \in B_x \right\}$. Then $C_x$ is an identity submatrix of $C$ of size $r$.
\end{lemma}
\begin{IEEEproof}
Note that there are $r$ columns in submatrix $C_x$ of $C$. Now we show that there are $r$ rows. Due to the fact that $({\cal X,A})$ is a BIBD with $\lambda =1$, the elements $\{x,A_{i}(j_{i} {\oplus_k} 1)\}\subset A_{i}$ only and does not lie in any other $A_{i'}$ for any $i'\neq i$. Thus the elements $A_{i}(j_{i} {\oplus_k} 1)\neq A_{i'}(j_{i'} {\oplus_k} 1)$ for any $i'\neq i$. Hence, there are $r$ rows in $C_x$. Let us consider the row indexed by $A_{i}(j_{i} {\oplus_k} 1)$. Again by the property that $\lambda =1$, it holds in $C_x$ that the only column index corresponding to which there is a $1$ in the row indexed by $A_{i}(j_{i} {\oplus_k} 1)$ is $A_{i}$ and no other column. Thus each row of $C_x$ has a single entry $1$ in some column. Now consider an arbitrary column of $C_x$, say indexed by $A_{i}$. Since $A_{i'}(j_{i'} {\oplus_k} 1)\in A_{i}$ only if $i=i'$, thus the column $A_{i}$ has a $1$ only in the row $A_{i}(j_{i} {\oplus_k} 1)$. Hence, $C_x$ is an identity submatrix of $C$ of size $r$.
\end{IEEEproof}



In next two lemmas we will prove that there is no overlap between the identity submatrices $C_x : x \in {\cal X}$ and that these identity submatrices will cover all the entries where $C(x,A) =1$ in matrix $C$.
\begin{lemma}
\label{no_overlap}
For distinct $x_1,x_2\in{\cal X}$, there is no $x\in {\cal X}$, $A\in{\cal A}$ with $C(x,A)=1$ such that $C(x,A)$ is covered by both $C_{x_1}$ and $C_{x_2}$, where $C_{x_1}$ and $C_{x_2}$ are as defined in Lemma \ref{Cx_identity}.
\end{lemma}
\begin{IEEEproof}
Suppose $C(x,A)=1$ is covered by both $C_{x_1},C_{x_2}$. By the definition of $C_{x_1}$ and $C_{x_2}$, it implies that $x_1, x_2 \in A$. Let the element $x_1$ and $x_2$ be present in $j$ and $j'$ position of $A$ where $j \neq j'$. Therefore, $x= A(j {\oplus_k} 1) =x= A(j' {\oplus_k} 1)$, by our construction. But $j \neq j'$. This gives a contradiction. Hence, there is no $C(x,A)=1$ which is covered by both $C_{x_1}$ and $C_{x_2}$. 
\end{IEEEproof}

\begin{lemma}
\label{C_cover}
The set of matrices $\{C_x :x\in {\cal X}\}$ forms an identity submatrix cover of $C$.
\end{lemma}
\begin{IEEEproof}
The total number of $1$'s in matrix $C$ is equal to the product of the number of $1$'s in each row and the number of rows, and thus equal to $rv$. For each $x\in {\cal X}$ (note that ${|\cal X|}=v$), there exists an identity submatrix $C_x$. From Lemma \ref{Cx_identity} and \ref{no_overlap}, we see that each identity submatrix is of size $r$ and no two such identity submatrices have overlaps.  These identity submatrices will cover $vr$ number of $1$'s in matrix $C$ which is equal to total number of $1$'s in $C$. Hence, $\{C_x :x\in {\cal X}\}$ forms an identity submatrix cover of $C$.
\end{IEEEproof}
We thus have the following theorem summarizing the caching scheme.
\begin{theorem}
\label{main}
The incidence matrix of a ($v,k,1$)- BIBD forms a $\left(K=v,F=\frac{v(v-1)}{k(k-1)},(1-\frac{M}{N})=\frac{k}{v}\right)$ caching matrix. Further there is a transmission scheme with rate $R = \frac{k(k-1)}{v-1}$.
\end{theorem}
\begin{IEEEproof}
The parameters of the caching matrix $C$ have already been defined. By Lemma \ref{Cx_identity}, \ref{no_overlap} and \ref{C_cover}, we have an identity submatrix cover consisting of $v$ identity submatrices. Hence in Theorem \ref{rate def}, $S=v$ and rate $R = \frac{S}{F}= \frac{k(k-1)}{v-1}$.

\end{IEEEproof}


\begin{example}
\label{ex6}
Consider the $(7,3,1)$-BIBD as given in Example \ref{ex_5}. We describe the identity submatrix $C_1$ (as per Lemma \ref{Cx_identity}) corresponding to element `$1$' in ${\cal X}$. Element `$1$' is present in blocks $\{127, 145, 136\}$. Then the columns of identity submatrix matrix $C_1$ are indexed by $127, 145, 136$ and rows are indexed by $2, 4, 3$ (since $2,4,3$ are the next elements present after `$1$' in blocks $127, 145, 136$ respectively) of matrix $C$. Similarly, the identity submatrix $C_6$ corresponding to element `$6$' in ${\cal X}$ has column indices and row indices as $136,467,256$ and $1,7,2$ respectively in matrix $C$. In this manner, the submatrices $C_i: i \in 
{\cal X}$, gives us a rate $1$ transmission scheme.
\end{example}

\section{Symmetric BIBD (with $\lambda =2$) based Coded Caching Scheme}
\label{Symm_BIBD}
Consider a symmetric ($v,k,2$)-BIBD $(\cal{X,A})$ with non-repeated blocks (i.e ${\cal A}$ is a set, not a multiset). Let the blocks in the BIBD be denoted as ${\cal A} = \left\{B_1,...,B_b\right\}$. Consider a set containing $kb$ items given by ${\cal B}= \left\{(B,i) : i\in B, B\in \cal A  \right\}$. Consider a binary matrix $M$ whose columns are indexed by elements of $\cal B$ and rows are indexed by elements of $\cal X$. The number of rows and the number of columns in matrix $M$ are $v$ and $kb=kv$ respectively. The matrix $M =(M(i,(B,j)))$ is defined by the rule

$M(i,(B,j)) =
  \begin{cases}
   1, & \text{if}\ i\in B, B\in {\cal A}, i\neq j \\
    0, & \text{otherwise}. \\
  \end{cases}$
  

\begin{remark}
Note that any $i\in {\cal X}$ occurs precisely in $k$ blocks (by (\ref{r_points})). Therefore, it follows that each row and column of matrix $M$ has a constant weight equal to $k(k-1)$ and $(k-1)$ respectively. Thus, binary caching matrix $M$ is a constant row and column weight matrix.
\end{remark}

 
Matrix $M$ gives a $(v,kv,\frac{k-1}{v})$ caching matrix. We now describe an identity submatrix of $M$.

\begin{lemma}
\label{symm_IM}
For each $i\in{\cal X}$ and $B\in {\cal A}$ such that $i \in B$, let $\left\{ B_j : j\in [k-1], B_j \neq B~\forall j \right\}$ be the $(k-1)$ blocks containing $i$ apart from $B$. Consider the submatrix $M_{i,B}$ of $M$ whose columns are indexed by $\left\{ (B_j,i) : j\in [k-1] \right\}$ and rows are indexed by $B \backslash i$. Then, $M_{i,B}$ is an identity submatrix of $M$ of size $(k-1)$.
\end{lemma}
\begin{IEEEproof}
Clearly the number of rows and columns are $(k-1)$. Fix some arbitrary row indexed by $x\in B\backslash i$. Note that $\{i,x\} \subset B$ and there is precisely one other $B_j \neq B$ such that $\{i,x\} \subset B_j$, since this is a BIBD with $\lambda =2$. Therefore the row $x$ contains a $1$ only in the column indexed by the unique $B_j$ such that $\{i,x\}\subset B_j$.

Fix some arbitrary column indexed by $(B_j,i)$. Thus $i\in B_j$. By (\ref{lambda}), any two blocks intersect at $\lambda=2$ points, so we must have some unique $x$ such that $\{x,i\}=B_j\cap B$. Therefore, the column indexed by $(B_j,i)$ contains $1$ only in the row indexed by $x$ and no other row of submatrix $M_{i,B}$. Hence, $M_{i,B}$ is an identity submatrix of $M$ of size $(k-1)$.

\end{IEEEproof}
In the next two lemmas we will prove that there is no overlap between the identity submatrices and that these identity submatrices will cover all the entries where $M(i,(B,j)) =1$ in matrix $M$.

\begin{lemma}
\label{symm_overlap}
Any $M(x,(B,j)) = 1$, will be covered by exactly one identity submatrix in the set $\{M_{i,{B'}} : \forall i\in B',\forall B'\in {\cal A} \}$ where the identity submatrices $M_{i,B'}$ are as defined in Lemma \ref{symm_IM}.
\end{lemma}
\begin{IEEEproof}
Suppose $M(x,(B,j)) = 1$ is covered by some identity submatrix $M_{i,{B'}}$ as defined in Lemma \ref{symm_IM}. Then by Lemma \ref{symm_IM}, it must be the case that $B\neq B', i=j$ and $\{x,j\}\in B'$. Since $\lambda=2$, there is precisely one block $B''$ apart from $B$ which contains $\{x,j\}$. Thus we can choose $B'=B''$ and this is the only choice for $B'$. Hence, the entry $M(x,(B,j)) = 1$ is covered only by the identity submatrix $M_{j,{B''}}$. This completes the proof.
\end{IEEEproof}
\begin{lemma}
\label{symm_main}
The set of matrices $\{M_{i,B} : \forall i \in B, \forall B \in {\cal A}\}$ forms an identity submatrix cover of $M$.
\end{lemma}
\begin{IEEEproof}
The total number of $1$'s in matrix $M$ is equal to the product of the number of $1$'s in each column and the number of columns, and thus equal to $(k-1)kb$. Each  $i\in {\cal X}$ (note that $|{\cal X}|=v)$, occurs in $k$ blocks since $(\cal {X,A})$ is a symmetric BIBD. Thus by Lemma \ref{symm_IM}, corresponding to each $i\in {\cal X}$, we get $k$ identity submatrices each of size $(k-1)$. By Lemma \ref{symm_overlap}, there is no overlap between any pair of identity submatrices among the set of all identity submatrices $\forall i \in {\cal X}$. Thus the number of $1$'s covered by all identity submatrices $= kv(k-1) = kb(k-1)$ ($\because  v=b$) which is equal to the total number of $1$'s in $M$. Hence, $\{M_{i,B} : \forall i \in B ,\forall B \in {\cal A} \}$ forms an identity submatrix cover of $M$.
\end{IEEEproof}
We now describe the following theorem summarizing the caching scheme.
\begin{theorem}
\label{symm_main_th}
The matrix $M$ of symmetric ($v,k,2$)-BIBD with non repeated blocks, forms a $\left(K=v,F=kv,(1-\frac{M}{N})=\frac{k-1}{v}\right)$ caching matrix. Further there is a transmission scheme with rate $R = 1$.
\end{theorem}

\begin{IEEEproof}
The parameters of the caching matrix $M$ have already been defined. By Lemma \ref{symm_IM}, \ref{symm_overlap} and \ref{symm_main}, we have an identity submatrix cover consisting of $v$ identity submatrices. Hence in Theorem \ref{rate def}, $S=v$ and rate $R = \frac{S}{F}=1$.

\end{IEEEproof}

\begin{example}
Consider the $(11,5,2)$-BIBD as given in Example \ref{ex2}. Let $M_{3,13467}$ be an identity submatrix corresponding to element `$3$' in ${\cal X}$ and block `$13467$' in ${\cal A}$. Element `$3$' is present in blocks $\{13467,1235a,23789,348ab,3569b\}$. Then by our construction, the columns of identity submatrix matrix $M_{3,13467}$ are indexed by  \{$(3,1235a),(3,23789),(3,348ab),(3,3569b)$\} and rows are indexed by $\{1,4,6,7\}$ (i.e $ \{1,3,4,6,7\}\backslash 3$) respectively of matrix $M$. In this manner we can obtain an identity submatrix cover of $M$ using the matrices $M_{i,B} : \forall i \in B, ~\forall B \in {\cal A}$, which gives us a transmission scheme with rate $=1$.
\end{example}

\section{$t$-design based Coded Caching Schemes}
\label{t-section}
We now describe two $t$-design based caching scheme. 
\subsection{Scheme-$1$}
\label{t_Scheme1}
Let $({\cal X},{\cal A})$ denote a $t$-$(v,k,1)$ design. Let the blocks in this $t$-design be denoted as ${\cal A}=\{B_1,...,B_b\}$ where $b=\frac{\binom{v}{t}}{\binom{k}{t}}$. We construct a binary matrix $T$ as follows.
Let the rows of $T$ be indexed by all the $(t-1)$-sized subsets of $\cal X$. Let the columns be indexed by $\{(y,B):y\in B, B\in{\cal A}\}$. The number rows in matrix $T$ is $\binom{v}{t-1}$. The number of columns in matrix $T$ is $bk=\frac{{\binom{v}{t}k}}{\binom{k}{t}}$. For some $D\in\binom{{\cal X}}{t-1}$, 
the matrix $T= (T(D,(y,B)))$ is defined by the rule,
\newline $T(D,(y,B)) =
  \begin{cases}
   1, & \text{if}\ D\cup\{y\}\subset B, |D\cup\{y\}|=t \\
    0, & \text{otherwise}. \\
  \end{cases}$
  
\begin{remark}  
The number of $1$'s in each row of $T$ is $\lambda_{t-1}\binom{k-t+1}{1}= (v-t+1)$. Also the number of $1$'s in each column is $\binom{k-1}{t-1}$. Hence, binary caching matrix $T$ is also a constant row and column weight matrix.
\end{remark}

Matrix $T$ gives a $\left(\binom{v}{t-1},\frac{{\binom{v}{t}k}}{\binom{k}{t}},\frac{{\binom{k}{t}(v-t+1)}}{{\binom{v}{t}}k}\right)$-caching matrix. In the next lemma we describe an identity submatrix of matrix $T$. Towards that end, we need to denote a few sets. For some $y \in {\cal X}$, define ${\cal B}_y \triangleq \{B \in {\cal A} : y \in B\}$, i.e the set of blocks containing $y$. By (\ref{s_points}), $|{\cal B}_y|=\lambda_1 = \frac{\binom{v-1}{t-1}}{\binom{k-1}{t-1}}$. Denote ${{\cal B}_y}$ by ${\cal B}_y=\{B_1,\hdots,B_{\lambda_1}\}$. For any $B_i$, denote by $\{D_{i,j} : j \in [\binom{k-1}{t-1}]\}$ the set $\binom{B_i\backslash y}{t-1}$, i.e the set of all $(t-1)$-sized subsets of $B_i\backslash y$.
\begin{lemma}
\label{t5_1}
For some $j\in[\binom{k-1}{t-1}] $ and $y \in {\cal X}$, consider the submatrix $T_{y,j}$ of $T$ whose rows are indexed by $\{D_{i,j} : \forall i \in [\lambda_1] \}$ ($D_{i,j}$ as defined above) and the columns are indexed by $\{(y, B_{i}): B_i \in {\cal B}_y\}$. Then $T_{y,j}$ is an identity submatrix of $T$ of size $\lambda_1$.   
\end{lemma}
\begin{IEEEproof}
Clearly, the number of columns in $T_{y,j}$ is $\lambda_1$. First note that the rows in $\{D_{i,j} : \forall i \in [\lambda_1]\}$ are all distinct, i.e $D_{i,j} \neq D_{{i'},j}$ for $i \neq {i'}$. If not, note that $D_{i,j} \cup y = D_{{i'},j} \cup y \in B_i \cap B_{i'}$. But this contradicts the fact that any $t$- sized subset of ${\cal X}$ occurs in only one block. Thus $|\{D_{i,j} : i \in [\lambda_1]\}|= \lambda_1$, and $T_{y,j}$ is a square matrix of size $\lambda_1$.

Now consider a row of $T_{y,j}$ indexed by $D_{i,j}$ for some particular $i \in [\lambda_1]$. Suppose a column indexed by $(y,B)$ for some $B \in {\cal B}_y$ has a $1$ in the row indexed by $D_{i,j}$. Then it means that $D_{i,j} \cup \{y\} \subset B$. But there is precisely one block $B$ such that $D_{i,j} \cup \{y\} \subset B$, which is precisely $B = B_i$ (as $\lambda=1$). Thus each row of $T_{y,j}$ has only one entry which is $1$.

Now, consider a column of $T_{y,j}$ indexed by $(y,B_i)$ for some $B_i \in {\cal B}_y$. Suppose for some $D_{{i'},j}$, the row indexed by $D_{{i'},j}$ has $1$ in the column indexed by $(y,B_i)$. Then it must be that $D_{{i'},j} \subset B_i \backslash y$ and hence $D_{{i'},j} \cup \{y\} \subset B_i$. Once again, because of the property of $t$-design with $\lambda=1$, we have that $i = {i'}$ (else $D_{{i'},j} \cup \{y\} \in B_i \cap B_{i'}$, which is a contradiction). Hence, each column of $T_{y,j}$ has precisely only one entry that is $1$. This proves the lemma.   
 \end{IEEEproof}
 
In the next two lemmas we will prove that there is no overlap between the identity submatrices and that these identity submatrices will cover all the entries where $T(D,(y,B)) =1$ in matrix $T$.
\begin{lemma}
\label{t5_2}
Any $T(D,(y,B)) =1$ such that $y\in {\cal X}, B\in{\cal A}, D\in \binom{\cal X}{t-1}$ will be covered by exactly one identity submatrix of $T$ (as defined in Lemma \ref{t5_1}).
\end{lemma}

\begin{IEEEproof}
Let $T(D,(y,B)) =1$ be covered by an identity submatrix $T_{{y'},j}$ of $T$.  As $T(D,(y,B)) =1$, we have that $D \cup \{ y\} \subset B$. By definition of $T_{{y'},j}$ in Lemma \ref{t5_1}, we must first have $y = {y'}$. Further it must be that $D \subset B_i \backslash y$ for some $B_i$ which contains $y$. Therefore, $\{y\} \cup D \subset B_i$, which means $B_i = B$ (as $\lambda=1$). Hence the unique transmission which covers $T(D,(y,B)) =1$ is $T_{y,j}$ where $j$ is such that $D=D_{i,j}$ is the unique $j^{th}$ set in $\binom{B \backslash y}{t-1}$.
\end{IEEEproof}

\begin{lemma}
\label{t5_3}
The set of matrices $\{T_{y,j}: \forall y\in {\cal X},~ \forall j\in [\binom{k-1}{t-1}]\}$ forms an identity submatrix cover of $T$.
\end{lemma}
\begin{IEEEproof}
The total number of $1$'s in $T$ is equal to the product of the number of $1$'s in each row and the number of rows, and thus equal to $(v-t+1)\binom{v}{t-1}=t \binom{v}{t}$. The size of each identity submatrix is $\frac{\binom{v-1}{t-1}}{\binom{k-1}{t-1}}$. Since we have an identity submatrix $T_{y,j}$ for each $j\in [\binom{k-1}{t-1}],~ y\in {\cal X}$, the number of identity submatrices $=\binom{k-1}{t-1}v$. Moreover, from Lemma \ref{t5_2}, there are no overlaps between the identity submatrices. Hence, the total number of $1$'s covered by all identity submatrices $= {\binom{k-1}{t-1}}{\frac{\binom{v-1}{t-1}v}{\binom{k-1}{t-1}}}=\binom{v-1}{t-1}v=t \binom{v}{t} $ which is equal to the total number of $1$'s in matrix $T$. Hence, $\{T_{y,j}: y\in {\cal X}, ~j\in [\binom{k-1}{t-1}]\}$ forms an identity submatrix cover of $T$.
\end{IEEEproof}
We thus have the below theorem summarizing the caching scheme.
\begin{theorem}
\label{T5_main}
The matrix $T$ of a $t$-($v,k,1$)-design forms a $\left(K=\binom{v}{t-1},F=\frac{{\binom{v}{t}k}}{\binom{k}{t}},(1-\frac{M}{N})=\frac{{\binom{k}{t}(v-t+1)}}{{\binom{v}{t}}k}\right)$-caching matrix. Further there is a transmission scheme with rate $R = \frac{{\binom{k-1}{t-1}}{\binom{k}{t}}v}{\binom{v}{t}k}$.
\end{theorem}

\begin{IEEEproof}
The parameters of the caching matrix $T$ have already been defined. By Lemma \ref{t5_1}, \ref{t5_2} and \ref{t5_3}, we have an identity submatrix cover consisting of $\binom{k-1}{t-1}v$ identity submatrices. Hence in Theorem \ref{rate def}, $S=\binom{k-1}{t-1}v$ and rate $R =\frac{S}{F}= \frac{{\binom{k-1}{t-1}}{\binom{k}{t}}v}{\binom{v}{t}k}$.


\end{IEEEproof}


\begin{example}
Consider the $3$-$(8,4,1)$ design as given in Example \ref{ex4}. We now describe some identity submatrices namely $T_{4,1},T_{4,2}$ and $T_{4,3}$. The set of blocks containing element `$4$' are denoted by ${\cal B}_4= \{3478,2468,1458,1234,3456,2457,1467\}$. It is clear from our construction that the rows of $T$ are indexed by $(t-1)=2$-sized subsets of ${\cal X}$. The identity submatrices $T_{4,1},T_{4,2}$ and $T_{4,3}$ have rows indexed by $\{37,26,15,12,35,25,16\},~ \{38,28,18,13,36,27,17\}$ and $\{78,68,58,23,56,57,67\}$ respectively and columns indexed by $\{(4,B_i): B_i\in {\cal B}_4 \}$ of matrix $T$. In this manner we can obtain an identity submatrix cover of $T$ using the matrices $T_{y,j}: \forall y \in {\cal X},~ \forall j \in [\binom{k-1}{t-1}]$, which gives a transmission scheme with rate $=\frac{3}{7}$.
\end{example}

\subsection{Scheme-$2$}
\label{t_Scheme2}
We now describe another caching matrix based scheme from a given Steiner system. Let $({\cal X},{\cal A})$ denote a $t$-$(v,k,1)$ design. Let the blocks in this $t$-design be denoted as ${\cal A}=\{B_1,...,B_b\}$ where $b=\dfrac{\binom{v}{t}}{\binom{k}{t}}$. For a block $B_j\in {\cal A}$, define ${\cal B}_j\triangleq\{(B_j,E): \forall E\in \binom{B_j}{t}\}$. Define ${\cal B}\triangleq \bigcup\limits_{j=1}^{b}{{\cal B}_j}$.

Consider a binary matrix $T$ whose columns are indexed by elements in ${\cal B}$ and rows are indexed by elements in ${\cal X}$. The number of rows and the number of columns in matrix $T$ are $v$ and $b\binom{k}{t}=\binom{v}{t}$ respectively. The matrix $T= (T(i,(B,E)))$ is defined by the rule,
\newline $T(i,(B,E)) =
  \begin{cases}
   1, & \text{if}\ i\in E, B\in {\cal A} \\
    0, & \text{otherwise}.\\
  \end{cases}$

\begin{remark}
Each row of $T$ has a constant weight equal to ${\lambda_1}\binom{k-1}{t-1}= \binom{v-1}{t-1}$ and each column has constant weight equal to $t$. Hence, binary caching matrix $T$ is a constant row and column weight matrix. 
\end{remark}

Matrix $T$ gives a $(v,\binom{v}{t},\frac{t}{v})$-caching matrix. Let ${\cal A}_D$ denote the set of blocks in ${\cal A}$ containing some $D\in \binom{{\cal X}}{t-1}$. We know that $|{\cal A}_D|={\lambda_{t-1}}=\frac{v-t+1}{k-t+1}$, by (\ref{s_points}). Thus we denote ${\cal A}_D$ as ${\cal A}_D=\{B_{D,1},\hdots,B_{D,{\lambda_{t-1}}}\}$. Note that in any $B_{D,j}$, the number of $E\in \binom{B_{D,j}}{t}$ such that $D\subset E$ is $(k-t+1)$. Denote the set $\{E\in \binom{B_{D,j}}{t}: D\subset E \}$ as $\{E_{D,j,i}: i\in[k-t+1] \}$. Now let ${\cal B}_D \triangleq \left\{(B_{D,j},E_{D,j,i})\in {\cal B}: \forall i\in [k-t+1], \forall j\in [\lambda_{t-1}] \right\}$. Thus $|{\cal B}_D|={\lambda_{t-1}}(k-t+1)=v-t+1$. In the next lemma, we describe a single identity submatrix of matrix $T$.
\begin{lemma}
\label{T_main}
For some $D\in \binom{{\cal X}}{t-1}$, consider a submatrix $T_D$ of $T$ whose columns are indexed by ${\cal B}_D$ and rows are indexed by $\{ E_{D,j,i}\backslash D : \forall i\in [k-t+1], \forall j\in [\lambda_{t-1}] \}$. Then $T_D$ is an identity submatrix of $T$ of size $(v-t+1)$.
\end{lemma}
\begin{IEEEproof}
Because of the fact that $({\cal X,A})$ is a $t$-design with $\lambda =1$, the set $E_{D,j,i}$ appears only in the block $B_{D,j}$ and no other block. Thus the elements $E_{D,j,i}\backslash D$, are all unique for each $j\in [\lambda_{t-1}], i\in [k-t+1]$. Hence, the matrix $T_D$ has $\lambda_{t-1}(k-t+1)=(v-t+1)$ rows. Therefore, $T_D$ is a square matrix of size $(v-t+1)$.

Consider any row of the submatrix $T_D$, indexed by some element $ E_{D,j,i}\backslash D \in {\cal X}$. Note that as $E_{D,j,i}\subset B_{D,j}$ and no other block, the row indexed by $(E_{D,j,i}\backslash D)$ has a $1$ only in the column of $T_D$ indexed by $(B_{D,j}, E_{D,j,i})$ and no other column of $T_D$.

Now consider any column of $T_D$, say indexed by $(B_{D,j},E_{D,j,i})$. Suppose some row indexed by $(E\backslash D)$ (for some $E \in \binom{B_{D,{j'}}}{t}$ for some $B_{D,{j'}}$) contains a $1$ in column $(B_{D,j},E_{D,j,i})$. Then it must be that $E \backslash D \in E_{D,j,i}$ (by definition of matrix $T$). But $E_{D,j,i}$ contains $D$ and hence $E = E_{D,j,i}$ (and thus $B_{D,{j'}}=B_{D,{j}}$). Thus the only row which contains 1 in column $(B_{D,j},E_{D,j,i})$ is $E_{D,j,i} \backslash D$. Hence, $T_D$ is an identity submatrix of $T$ of size $(v-t+1)$. 
\end{IEEEproof}

In the next two lemmas, we will prove that there is no overlap between the identity submatrices and that these identity submatrices will cover all the entries where $T(i,(B,E))=1$ in matrix $T$.
 \begin{lemma}
\label{T_overlap}
For distinct $D_1,D_2 \in \binom{{\cal X}}{t-1}$, there is no $T(x,(B,E))=1$ where $x \in {\cal X},~E \subset B,~ B \in {\cal A}$ such that $T(x,(B,E))$ is covered by both identity submatrices $T_{D_1}$ and $T_{D_2}$, where $T_{D_1}$ and $T_{D_2}$ are as defined in Lemma \ref{T_main}.
\end{lemma}
\begin{IEEEproof}
 Suppose $T(x,(B,E))=1$ is covered by both $T_{D_1}$ and $T_{D_2}$. From Lemma \ref{T_main}, row $x= E \backslash D_1 = E \backslash D_2 $. But, $D_1 \neq D_2$. This gives a contradiction. Hence, there is no $T(x,(B,E))=1$ such that $T(x,(B,E))$ is covered by both identity submatrices $T_{D_1}$ and $T_{D_2}$.
\end{IEEEproof}

\begin{lemma}
\label{T_cover}
The set of matrices $\{T_D$ : $D \in \binom{\cal X}{t-1}\}$, forms an identity submatrix cover of $T$.
\end{lemma}
\begin{IEEEproof}
The total number of $1$'s in $T$ is equal to the product of the number of $1$'s in each column and the number of columns, and thus is equal to $t \binom{v}{t}$. By Lemma \ref{T_main}, corresponding to each $D \in \binom{{\cal X}}{t-1}$, we get an identity submatrix, so $|\{T_D$ : $D \in \binom{\cal X}{t-1}\}|=\binom{v}{t-1}$. Each such identity submatrix is of size $(v-t+1)$. By Lemma \ref{T_overlap}, there is no overlap between any pair of such identity submatrices. Thus the total number of $1$'s covered by all the identity submatrices $=(v-t+1) \binom{v}{t-1}=t \binom{v}{t}$ which is equal to the total number of $1$'s in $T$.  Hence, $\{T_D$ : $D \in \binom{\cal X}{t-1}\}$ forms an identity submatrix cover of $T$.
\end{IEEEproof}
We now have the following theorem summarizing the caching scheme.
\begin{theorem}
\label{T_main_th}
The matrix $T$ of a $t$-($v,k,1$)-design, forms a $\left(K=v,F=\binom{v}{t},(1-\frac{M}{N})=\frac{t}{v}\right)$ caching matrix. Further there is a transmission scheme with rate $R = \frac{t}{v-t+1}$.
\end{theorem}

\begin{IEEEproof}
The parameters of the caching matrix $T$ have already been defined. By Lemma \ref{T_main}, \ref{T_overlap} and \ref{T_cover}, we have an identity submatrix cover consisting of $\binom{v}{t-1}$ identity submatrices. Hence in Theorem \ref{rate def}, $S=\binom{v}{t-1}$ and rate $R =\frac{S}{F}=\frac{\binom{v}{t-1}}{\binom{v}{t}}=\frac{t}{v-t+1}$.


\end{IEEEproof}

\begin{example}
Consider the $3$-$(8,4,1)$ design as given in Example \ref{ex4}. We now describe some identity submatrix namely $T_{\{1,2\}}$. The set of blocks containing $D=\{1,2\}$ are denoted by ${\cal A}_{\{1,2\}}= \{1256,1234,1278\}$ (as $\lambda_{t-1}=3$). So
${\cal B}_{\{1,2\}} = \{(1256,125),(1256,126),(1234,123),
(1234,124),$\newline $(1278,127),(1278,128)\}$. Then the identity submatrix $T_{\{1,2\}}$ has columns indexed by ${\cal B}_{\{1,2\}}$ and rows indexed by $\{5,6,3,4,7,8\}$ of matrix $T$. In this way we can obtain an identity submatrix cover of $T$ using the matrices $T_D: \forall D \in \binom{{\cal X}}{2}$, which gives a transmission scheme with rate $=\frac{1}{2}$.

\end{example}

\begin{remark}
For given $K,M,N$, such that $K(1-\frac{M}{N})$ is an integer, if we take ${\cal X}=[K]$ and ${\cal A}$ as all the $K(1-\frac{M}{N})$ sized subsets of ${\cal X}$ we get a $t$-design with parameters $t=k=K(1-\frac{M}{N})$, $v=|{\cal X}|=K$ and $\lambda =1$. For Scheme-2, this $t$-design will give rate $R=\dfrac{t}{v-t+1}=\dfrac{K(1-\frac{M}{N})}{1+\frac{KM}{N}}$, which is same as the rate obtained in \cite{MaN}. Indeed the other parameters also match with the scheme in \cite{MaN}.
\end{remark}

\section{Transversal Design based Coded Caching Scheme}
\label{transvelsal}
Consider a TD($k,n$) transversal design $ ({\cal X}, {\cal G}, {\cal B})$ such that $k\geq n$. We order the group in some particular way. Thus, ${\cal G}=\{G_i : i \in \{0,1,\hdots,k-1\}\}$ where $|G_i|= n$, denotes the groups. Note that $|{\cal X}|=kn$. We see that by (TD$4$), for $\lambda =1$ each pair of elements from ${\cal X}$ is contained either in exactly one group in ${\cal G}$ or in one block in ${\cal B}$, but not both. Thus any two elements in a block must be from different groups. Therefore, for any block $A\in {\cal B}~(|A|=k)$, there is precisely one element  $A({G_i})$, such that $A({G_i}) \in B \cap G_i$. Thus, the elements in block $A\in {\cal B}$ cab be ordered as $\{ A(G_0),\hdots,A(G_{k-1})\}$ where element $A(G_i), i \in \{0,1,\hdots,k-1\} $ is from group $G_i$. 

Let $C$ be the transpose of the incidence matrix of TD$(k,n)$. The number of rows and the number of columns in matrix $C$ are $n^2$ and $kn$ respectively.

\begin{remark}
 Each row and column of matrix $C$ has a constant weight equal to $k$ and $n$ respectively. Hence, matrix $C$ is a constant row and column weight matrix.
\end{remark}

Matrix $C$ will give a ($n^{2},kn,\frac{1}{n}$)-caching matrix. Recall that each $x \in {\cal X}$ occurs in precisely $n$ blocks of ${\cal B}$, by the property of TD$(k,n)$. We now describe the lemma which gives an identity submatrix of matrix $C$. 


\begin{lemma}
\label{td_Cx_identity}
Define by ${\cal B}_x$ the set of blocks which contain $x$ and denote ${{\cal B}_x} = \{A_{1},\hdots,A_{n}\}$ where $x=A_{i}(G_{j_i}) : {j_i}\in \{0,1,\hdots,k-1\}$ i.e. $x$ is the ${j_i}$ element in block $A_i$. Consider the submatrix $C_x$ whose rows are indexed by ${\cal B}_x$ and columns are indexed by $\left\{A_{i}(G_{j_{i}{\oplus_k}1}): A_{i}\in {\cal B}_x  \right\}$. Then $C_x$ is an identity submatrix of $C$ of size $n$.
\end{lemma}
\begin{IEEEproof}
Note that there are $n$ rows in submatrix $C_x$ of $C$. Now we show that there are $n$ columns. By the property of TD with $\lambda =1$, the elements $\{x,A_{i}(G_{j_{i}{\oplus_k}1}) \}\subset A_{i}$ only and does not lie in any other $A_{i'}$ for any $i'\neq i$. Thus the elements $A_{i}(G_{j_{i}{\oplus_k}1})\neq A_{i'}(G_{j_{i'}{\oplus_k}1})$ for any $i'\neq i$. Hence, there are $n$ columns in $C_x$.

Fix any arbitrary row indexed by $A_{i} \in {\cal B}_x$. Suppose this row has entry $1$ in two distinct columns indexed by $y$ and ${y'}$ of submatrix $C_x$ where $y,{y'} \in {\cal X}$. This means that $y,{y'} \in A_{i}$ and thus $\{x,y,y'\} \subseteq A_i$, which means $\{x,y\} \nsubseteq A_i'$ and $\{x,y'\} \nsubseteq A_i'$ for any $i\neq i'$.	 This means both $y,{y'} \in G_{j_{i}{\oplus_k}1}$. By the property (TD4) for $\lambda=1$, this is not possible. Hence, each row of submatrix $C_x$ has entry $1$ in only one column.

Suppose the column $A_{i}(G_{j_{i}{\oplus_k}1})$ has $1$ in two rows, say $A_{i}$ and $A_{i'}$ such that $i \neq {i'}$. Then the set $\left\{x,A_{i}(G_{j_{i}{\oplus_k}1}) \right\}$ occurs in both blocks $A_{i}$ and $A_{i'}$. Thus we have a contradiction with the property of Transversal design for $\lambda=1$ i.e. the pair $\{x,A_{i}(G_{j_{i}{\oplus_k}1}) \}\in {\cal X}$ is contained in either one group or in exactly one block, but not both. Hence, $C_x$ is an identity submatrix of $C$ of size $n$.  
\end{IEEEproof}

\begin{lemma}
\label{TD_main}
For distinct $x_1,x_2\in{\cal X} $, there is no $x\in {\cal X}$, $A \in {\cal B}$ with $C(A,x)=1$ such that $C(A,x)$ is covered by both $C_{x_1}$ and $C_{x_2}$. Further, the set of matrices $\{C_x :x\in {\cal X}\}$ forms an identity submatrix cover of $C$ of size $n$.
\end{lemma}
\begin{IEEEproof}
The first part of the proof follows the same arguments as Lemma \ref{no_overlap}. The total number of $1$'s in matrix $C$ is equal to the product of number of $1$'s in each row and the number of rows, and thus equal to $kn^2$. For each $x\in {\cal X}$ (note that $|{\cal X}|=kn)$, there exists an identity submatrix $C_x$. From the first part of the proof and Lemma \ref{td_Cx_identity}, we see that each identity submatrix is of size $n$ with no overlaps. These identity submatrices will cover $kn^{2}$ number of $1$'s in matrix $C$ which is equal to total number of $1$'s in $C$. Hence, $\{C_x :x\in {\cal X}\}$ forms an identity submatrix cover of $C$ of size $n$. 
\end{IEEEproof}
We thus have the following theorem summarizing the caching scheme.
\begin{theorem}
\label{td_main_th}
The transpose of the incidence matrix of a TD-($k,n$) forms a $\left(K=n^{2},F=kn,(1-\frac{M}{N})=\frac{1}{n}\right)$ caching matrix. Further there is a transmission scheme with rate $R = 1$.
\end{theorem}

\begin{IEEEproof}
The parameters of the caching matrix $C$ have already been defined. By Lemma \ref{td_Cx_identity} and \ref{TD_main}, we have an identity submatrix cover consisting of $kn$ identity submatrices. Hence in Theorem \ref{rate def}, $S=kn$ and rate $R  =\frac{S}{F}=1$.


\end{IEEEproof}

\begin{example}
Consider the TD$(4,3)$ as given in example \ref{ex5}. We describe some identity submatrix $C_2$ corresponding to element `$2$' in ${\cal X}$. Element `$2$' is present in blocks $\{\{2,4,9,11\},\{2,5,7,12\},\{2,6,8,10\}\}$. Hence, the rows of the identity submatrix matrix $C_2$ are indexed by $\{2,4,9,11\}, \{2,5,7,12\}, \{2,6,8,10\}$ and columns are indexed by $4,5,6$ (since $4,5,6$ are the next elements present after `$2$' in blocks $\{2,4,9,11\}, \{2,5,7,12\}, \{2,6,8,10\}$ respectively) of matrix $C$. In this manner, the submatrices $C_i: i \in 
{\cal X}$, gives us a rate $1$ transmission scheme.
\end{example}


\section{Discussion}
\label{discussion}
     A prior work, \cite{RM}, had already initiated the study of connections between coded caching and combinatorial designs by looking at caching schemes constructed via resolvable designs, for which the authors provide a coding theoretic construction. 
     However the approach of \cite{RM} is different from the approach we take here. In particular, in \cite{RM}, the property of a combinatorial design being `resolvable' was used to design the delivery scheme. In our work, there is no explicit requirement for resolvability of the design. Rather, the fundamental structure of the designs themselves give raise to techniques using which we can construct delivery schemes.
    Inspite of having small rates and subpacketization levels, the schemes constructed here still suffer from the drawback of requiring large local cache sizes, which is impractical. This can possibly be remedied by looking at designs with a higher value of $\lambda$. However it is so far unclear as to how we can design good delivery schemes for such cases. Nevertheless, we believe that this line of questioning will be worthwhile to explore.
\bibliography{IEEEabrv,root.bib}
\bibliographystyle{ieeetr}
\end{document}